\documentclass[aps,pre,reprint]{revtex4-1}

\usepackage{amsmath}

\usepackage{graphicx}
\usepackage{caption}
\usepackage{subcaption}
\usepackage[varg]{txfonts}

\newcommand{\degree}{\ensuremath{^\circ}}
\newcommand{\degC}{\ensuremath{\degree\mathrm{C}}}
\newcommand{\Eqref}[1]{Equation~\ref{#1}}
\newcommand{\Figref}[1]{Figure~\ref{#1}}

\newcommand{\secref}[1]{Section~\ref{#1}}
\newcommand{\Tabref}[1]{Table~\ref{#1}}
\newcommand{\tenpow}[1]{\ensuremath{\times10^{#1}}}

\newcommand{\kT}{\ensuremath{k\,T}}
\newcommand{\Rey}{\ensuremath{\mathrm{Re}}}

\newcommand{\suppl}[1]{%
  \ifx&#1&%
  (see Supplementary Material)%
  \else
  (see Supplementary Material {#1})%
  \fi
}

\begin{document}


\title[Rapid Spontaneous Lipid Assembly]{Rapid Spontaneous Assembly of
  Single Component Liposomes}

\author{P. Sunthar}
\email{P.Sunthar@iitb.ac.in}
\author{Sopan M. Phapal}
\affiliation{Department of Chemical Engineering, Indian Institute of Technology
  Bombay (IITB), Powai, Mumbai 400076, India.}

\begin{abstract}

  We present a mechanism and show two variants of a method where the
  average diameter of spontaneously (barrier-free) assembled single
  component unilamellar liposomes is intrinsic, in agreement with
  Helfrich's theory. It depends only on the temperature and the lipid
  type, eliminating kinetic effects or external forcing normally
  observed.  This provides the first pure system to study the
  self-assembly of vesicle forming components, and with a natural
  length scale it may have an implication for vesicle size selection
  under pre-biotic conditions.
\end{abstract}






\maketitle


Current hypotheses for the formation of enclosing membranes of
protocells (pre-biotic vesicles), are mainly on possible surfactants,
their origins or synthesis routes, autocatalysis, and their chemical
evolution
\cite{Deamer1985,Luisi1999,Szostak2001,Rasmussenetal2009}. While these
chemistries are the most crucial elements in solving the puzzle of
origin of cellular life, the physics of self-assembly has not seen any
significant attention in this context. Helfrich¡¯s free energy model
\citep{Helfrich1974115} provides the simplest description of vesicle
self-assembly from solution, which has a natural length scale for its
average diameter. We show how this can be achieved through a truly
spontaneous process (barrier free, unlike the externally driven
systems known thus far) for the simplest bilayer forming system
(single component lipid). If having a reproducible diameter is
important to protocells, then Helfrich¡¯s theory is the simplest
physical model that provides one, and the stationary phase
interdiffusion (SPI) mechanism proposed here is one way to attain it
spontaneously.  This is applicable to any vesicle forming system, with
suitably chosen solvents.

A suspension of unilamellar vesicles is kinetically trapped in high
energy states.  The vesicle diameter is usually determined by external
parameters from the sample preparation process which leads to one of
these energy traps \suppl{\secref{sec:vesthermo} for the
  thermodynamics of vesicles} \citep{Lasic2001, Grafmulleretal2007}.
Nevertheless, there is \emph{one} characteristic diameter, for
vesicles spontaneously assembled from a solution phase of lipids,
obtained from a energy balance on a disc micelle aggregate (bilayer
disc membrane)
\cite{Helfrich1974115,Fromherz1983,Sackmann1995a,Boal2012}. This
diameter $D$ is given by
\begin{equation}
  \label{eq:dint}
  D \approx 4 \, \frac{\left( 2\kappa + \bar{\kappa} \right)}{\gamma} 
\end{equation}
where $\kappa$ is the elastic bending modulus of the membrane,
$\bar{\kappa}$ is its Gaussian curvature modulus \citep{Huetal2012},
and $\gamma$ is the line (edge) tension of the membrane with the
surrounding solvent.  This diameter has also been known as the
``minimum'' diameter, obtainable by intense sonication
\citep{Helfrich1974115,Maulucci2005}.  However, we choose to call it
as an \emph{intrinsic size} of the vesicle, as $\kappa$,
$\bar{\kappa}$, and $\gamma$ are thermodynamic parameters of the
vesicle forming amphiphile in the surrounding solvent \suppl{
  \secref{sec:min-int} for an elaboration on this point}.  For
phosphotidylcholine lipid vesicles in water this theoretical diameter
can be estimated to be 13~nm \suppl{\secref{sec:dint} for the
  calculation} \cite{LyLongo2004,Sanaiotti2010,Karatekin2003}.
Despite being a natural length scale, this diameter has not been
observed by a spontaneous process starting from a solution phase of a
single component lipid system.

Strong hydrophobic interaction of double tailed surfactants and the
high energy state of the intrinsic diameter are the chief reasons
behind the difficulty of observing such a spontaneous assembly
\cite{Laughlin1997,Lasic2001}.  The small critical aggregation
concentrations of phospholipids (typically less than a nM), does not
permit efficient dispersion and mixing of the surfactants.
Conventional methods therefore require external energy (such as
sonication or extrusion) to be dissipated in the system
\cite{Maulucci2005}, using multi-component systems (such as
catanionic, lecithin+bile, etc.)
\cite{Hauser1989,Gabriel1984,Kaler1989,Leng2002,Segota2006}, or by
inducing a spontaneous curvature through certain molecules
\cite{Safran1990,Lasic1993liposomes} (for multi-component systems,
spontaneous formation occurs when kinetic effects of other surfactants
plays a role in modifying the edge energy \cite{Leng2002}).

We conceptualise a mechanism to observe the spontaneous assembly.
Consider a uniformly dispersed solution of bilayer forming molecules.
If the bulk (good) solvent is suddenly replaced by an aqueous (poor)
solvent, the changed conditions favour aggregation of the molecules to
form circular discs (bilayer membranes or bicelles). The system then
gets trapped in its intrinsic size as the discs grow. Such a situation
of ``magically'' replacing the surroundings can be easily achieved in
computer simulations of molecules
\cite{Markvoortetal2005,Noguchi2006,Huetal2012}, but the challenge is to realise
it in reality.

\begin{figure}[tbp]
  \centering
  \includegraphics[width=0.8\linewidth]{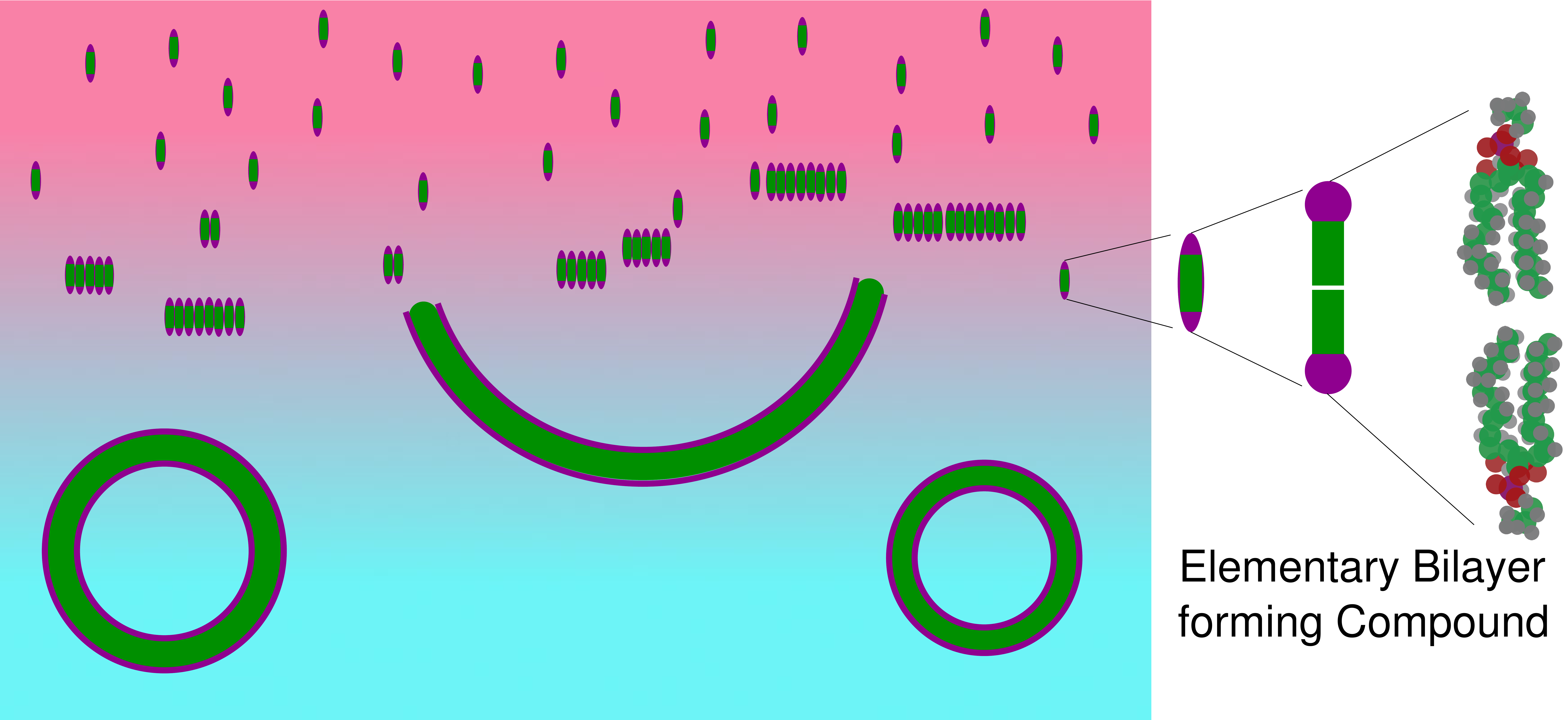}
  \caption{\textbf{Mechanism of spontaneous assembly in a nearly
      two-dimensional plane} Here the elementary bilayer forming
    compound (lipids for example) is generically represented
    schematically by a hydrophobic body (green) and two polar ends
    (purple). There are two solvents that are miscible with each other
    and the elementary compound is highly soluble in one (top; magenta)
    and poorly soluble in the other (bottom; cyan). At the interface
    as the solvents interdiffuse, the environment of the compound is
    rapidly replaced by the poor solvent which favours self-assembly
    and formation of closed vesicles. \emph{Note}: The representation
    of two lipids together as a single unit in the good solvent is
    only for illustrating a generic principle; in reality the lipid
    coordinates will be different.}
  \label{fig:mech}
\end{figure}

Instead of a bulk replacement, the solvents can be diffusively
replaced across an interface, provided they are miscible.  When the
interdiffusion coefficient of the solvents is much large compared to
that of the amphiphile molecule in the original solvent (lipids
lengths are about 10 times the diameter of the water molecules, and
have a correspondingly smaller diffusivity), these molecules will be
localised in space while the original solvent is replaced by the
aqueous solvent by interdiffusion, as illustrated in the schematic
shown in \Figref{fig:mech}.  This results in a condition analogous to
the sudden bulk replacement discussed above, albeit in a nearly
two-dimensional region.  This region of the solution is essentially
subjected to a ``quench'' from a dilute or ``gaseous'' phase (of the
amphiphiles) to a spinodal region leading to a bulk phase separation
to the bilayer phase. The assembly of the free surfactants happens
spontaneously without any barrier to form discs.  These discs, above
the critical diameter given by \Eqref{eq:dint} again transition,
without any barriers, to a lower energy state of a vesicle
\suppl{\secref{sec:vesthermo}}.  The size
of the vesicle would be intrinsic, but at a value determined by the
thermodynamics of the lipid in the mixture of solvents.

To achieve this mode of assembly, three components are essential:
bilayer forming amphiphile (lipid or a block co-polymer), a good
solvent and a poor solvent which are inter-miscible.  We found a
classical system---phosphotidylcholine (PC) lipid, ethanol, and
water---well suited to capture the phenomenon described. Here, the
lipids dissolved in ethanol is brought into contact with water across
a stationary interface.

Owing to the high miscibility of water and ethanol the interfaces need
to be brought into contact in a \emph{nearly stationary} manner,
taking care to avoid any convective (advective) mixing, that can
influence the energetics of self-assembly \cite{Phapal2013} \suppl{for
  a description of the setup and the method}. Achieving the stationary
contact sets this apart from the broad class of ethanol injection
methods using the same components
\cite{Phapal2013,Batzri1973,Jahn20042674}.  As the interdiffusion
proceeds, a turbid front is seen to move upwards, indicating the
formation of liposomes \suppl{for a detailed description and a video}.

\begin{figure*}[tbp]
  \begin{subfigure}{0.3\linewidth}
    \centering
    \includegraphics[width=\textwidth]{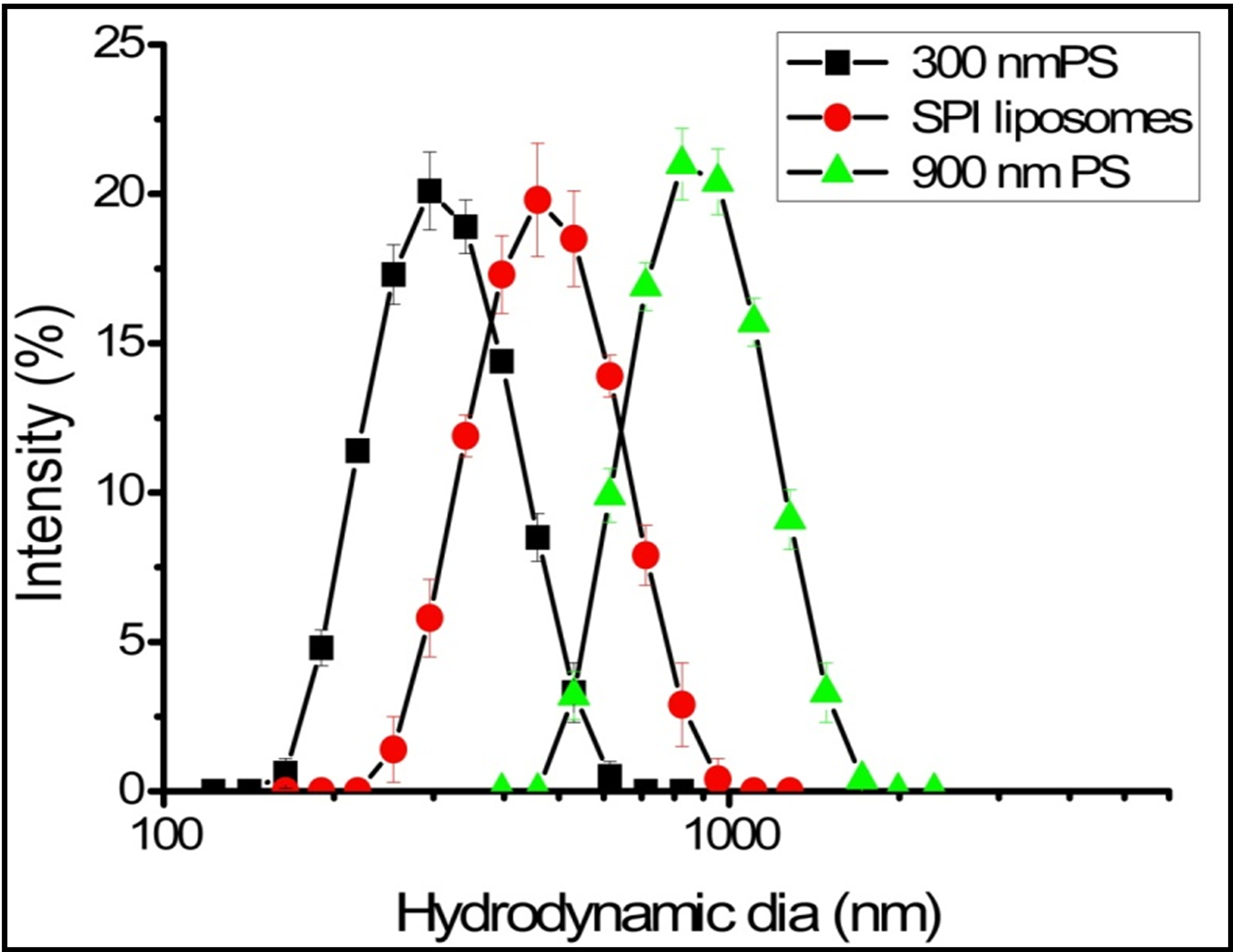}
    \caption{DLS}
    \label{fig:dls}
  \end{subfigure}
  \begin{subfigure}{0.2\linewidth}
    \centering
    \includegraphics[width=\linewidth]{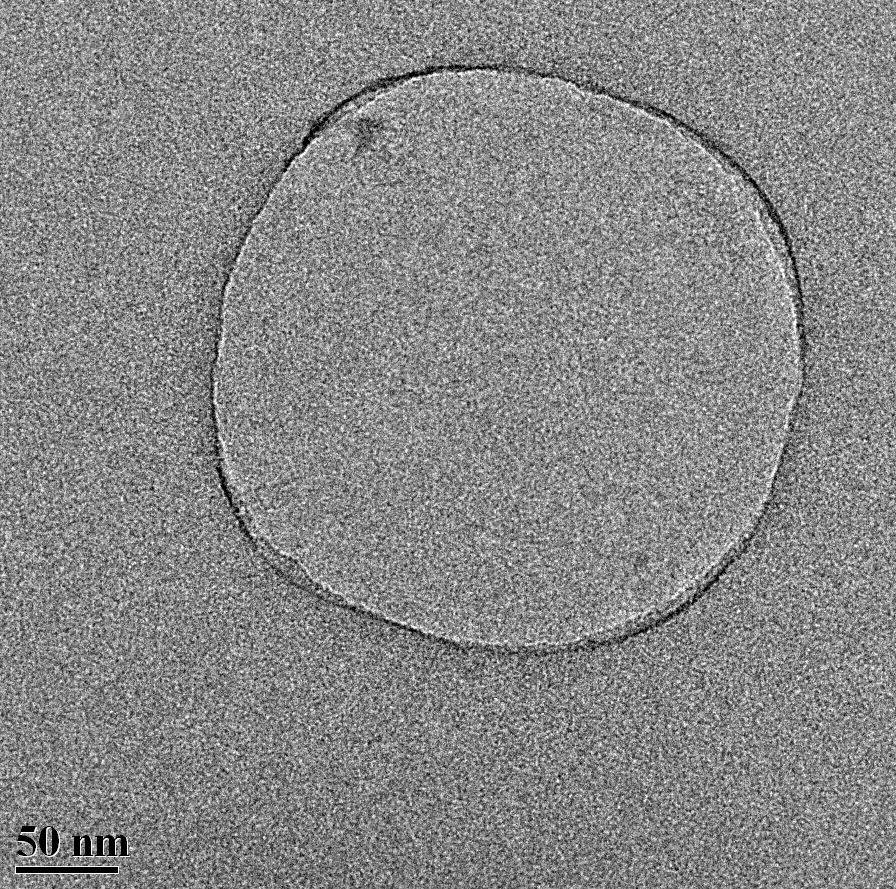}
    \caption{Dimyristoyl PC (DMPC)}
    \label{fig:hrtem-dmpc}
  \end{subfigure}
  \begin{subfigure}{0.2\linewidth}
    \centering
    \includegraphics[width=\linewidth]{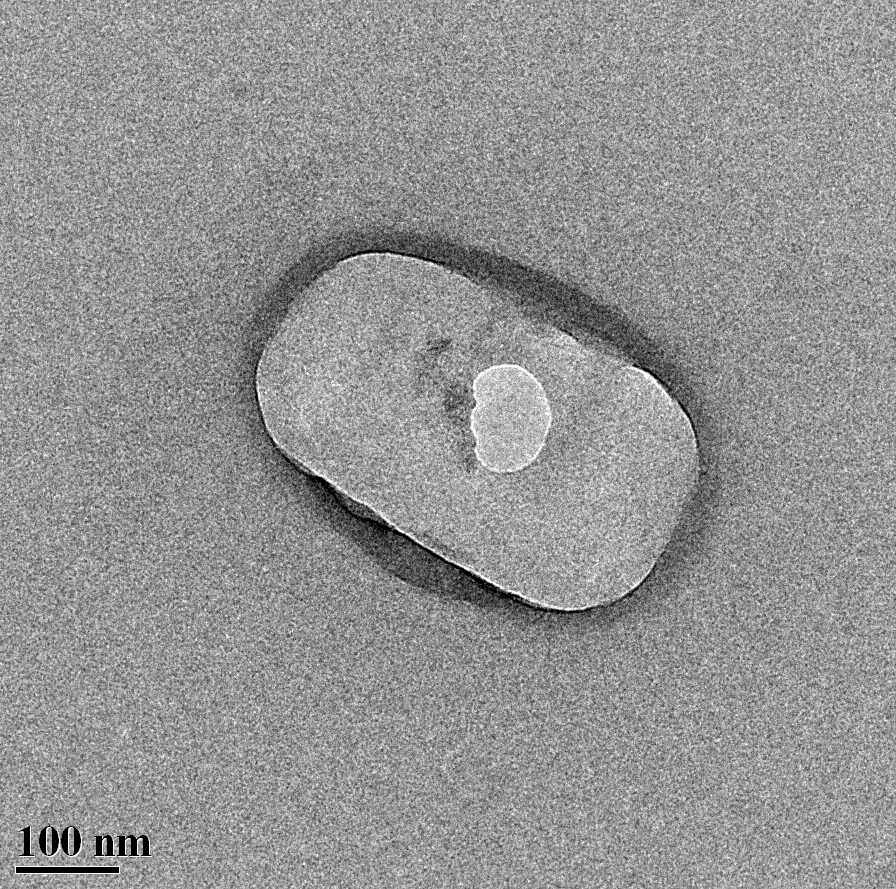}
    \caption{Dipalmitoyl PC (DPPC)}
    \label{fig:hrtem-dppc}
  \end{subfigure}
  \begin{subfigure}{0.2\linewidth}
    \centering
    \includegraphics[width=\linewidth]{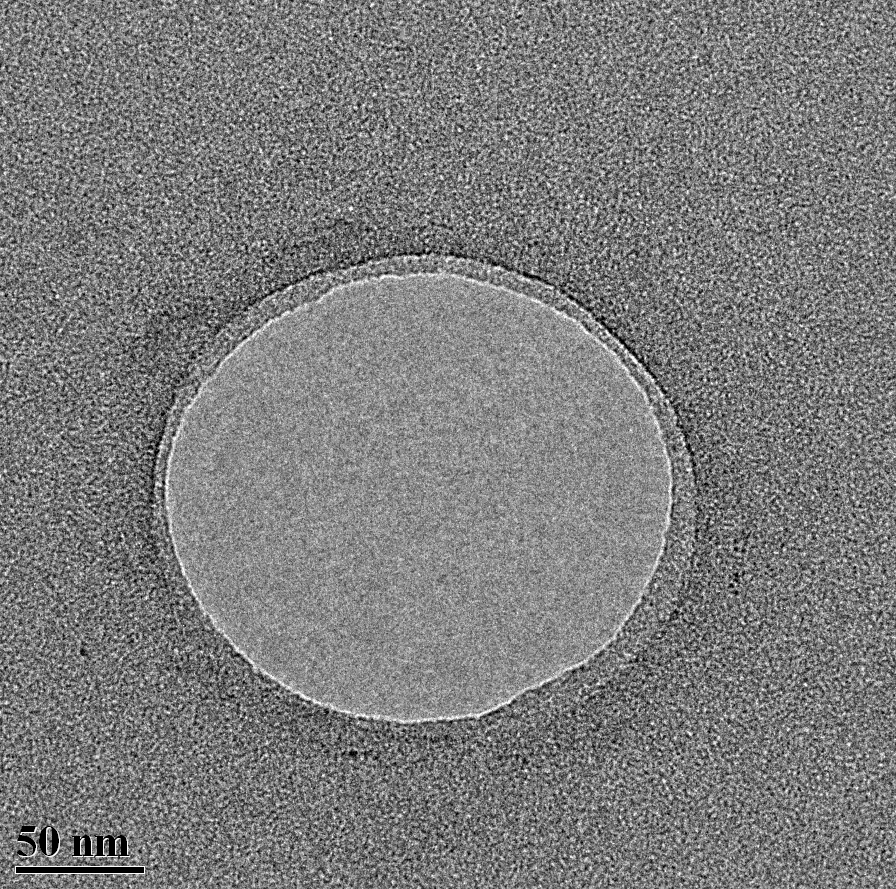}
    \caption{Soy PC}
    \label{fig:hrtem-spc}
  \end{subfigure}
  \caption[Liposome characterization]{\textbf{Liposome
      characterisation} (a) Monodisperse size distribution of DMPC
    liposomes as intensity of scattered light obtained by dynamic
    light scattering (DLS), shown in comparison with polystyrene
    standards of monodisperse population.   Here, the average size of
    liposomes, determined from the intensity distribution, is
    530$\pm$25~nm. (b-d) Negative staining micrographs from HRTEM showing
    a large unilamellar liposomes from various lipids.}
  \label{fig:character}
\end{figure*}

The liposome suspension obtained shows a monodisperse population with
a mean hydrodynamic diameter equal to 530$\pm$25~nm for the
dimyristoyl phosphotidylcholine (DMPC) lipid as seen in
\Figref{fig:character}(a). High resolution transmission electron
microscope (HRTEM) images under negative staining show that the
liposomes are unilamellar for various types of lipids used
(\Figref{fig:character}(b-d)).  An independent estimate of the
intrinsic diameter using \Eqref{eq:dint} can be obtained. In the
presence of ethanol we find this to be about 300~nm for the DMPC
liposome, which is a reasonable order of magnitude estimate of the
experimentally observed hydrodynamic diameter of 530~nm, given the
approximate estimates involved \suppl{ \secref{sec:dint} for the
  calculations}.

\begin{figure*}[tbp]
  \begin{subfigure}{0.48\textwidth}
    \centering
    \includegraphics[width=\linewidth]{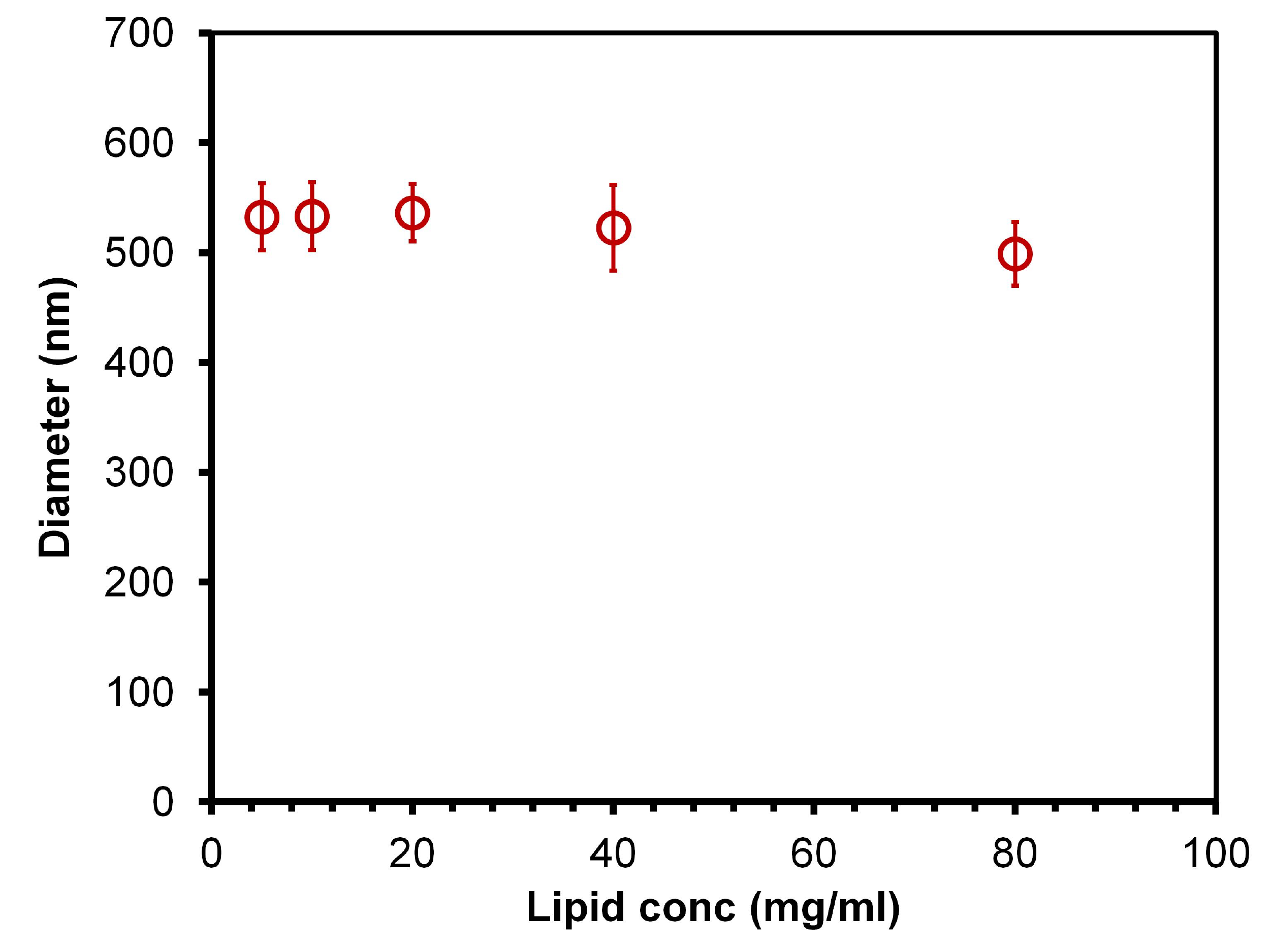}
    \caption{Effect of lipid concentration}
    \label{fig:lipidconc}
  \end{subfigure}
  \begin{subfigure}{0.48\textwidth}
    \centering
    \includegraphics[width=\linewidth]{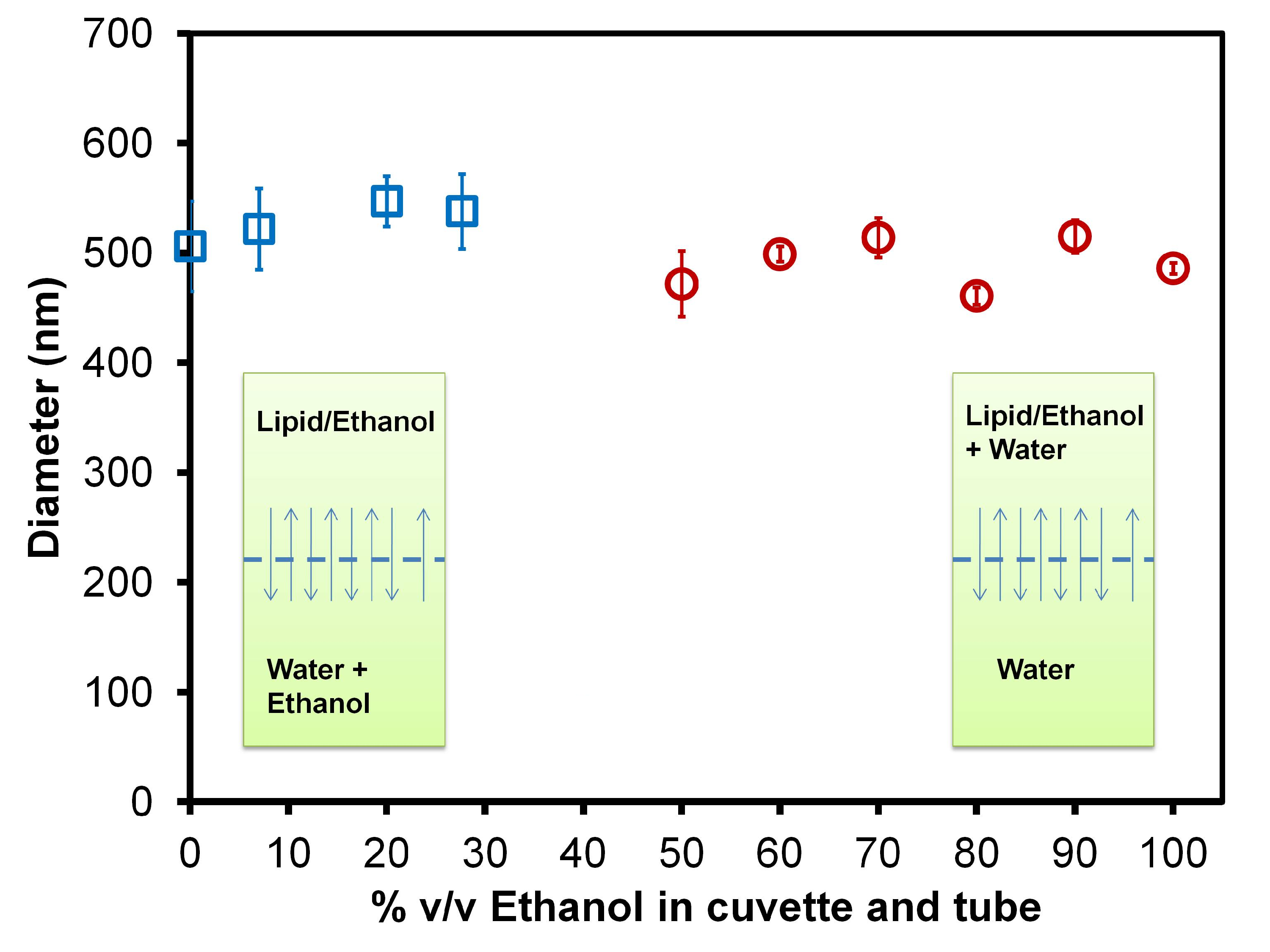}
    \caption{Effect of ethanol/water ratio}
    \label{fig:etohconc}
  \end{subfigure}
  \caption{ \textbf{Liposome size is intrinsic}
    (a) There is no effect of the lipid (DMPC) concentration initially
    dissolved in ethanol on the diameter of the liposomes formed.  (b)
    Evidence that shows that concentration of ethanol or its gradients
    across the interface does not influence the diameter of
    liposomes. The points on the left (squares in blue) were obtained
    by the pre-mixing ethanol to the aqueous phase in the cuvette,
    whereas those on the right (circles in red) were obtained by
    pre-mixing water in the lipid+ethanol solution in the syringe
    tube.  Except for a small region in ethanol fraction (between
    25--50\% v/v), where no liposomes are stably formed, the diameter
    of the liposomes is nearly constant at 510 $\pm$ 50~nm.}
  \label{fig:conceff}
\end{figure*}

That the diameter is intrinsic is reinforced by two crucial
observations. (i) The concentration of lipids does not have an
influence on the diameter of vesicles as seen in
\Figref{fig:lipidconc}.  Higher concentration only leads to higher
number density of vesicles of the same average size, as evidenced by
the same number counts from scattering measurements achieved at a
higher dilution and visually by a more turbid suspension \suppl{
  \secref{sec:conc} for calculations supporting this claim}.
(ii) the proportion of ethanol and water on either side of the
interface does not influence the size so long as the vesicles are
formed, as seen in \Figref{fig:etohconc}.  We infer from this that the
lipid assembly happens when the bulk concentration of ethanol in the
interfacial region is between 25 and 50\% (v/v).

Another important inference from \Figref{fig:conceff} is that in this
method, kinetic effects do not seem to play a role in determining the
average size of the vesicles. Otherwise the
concentration of lipids or the magnitude of the gradient of ethanol
across the interface would have altered the size, as may be expected
in a detergent depletion method \citep{Leng2003}. 

\begin{figure*}[tbp]
  \begin{subfigure}{0.48\textwidth}
    \centering
    \includegraphics[width=\linewidth]{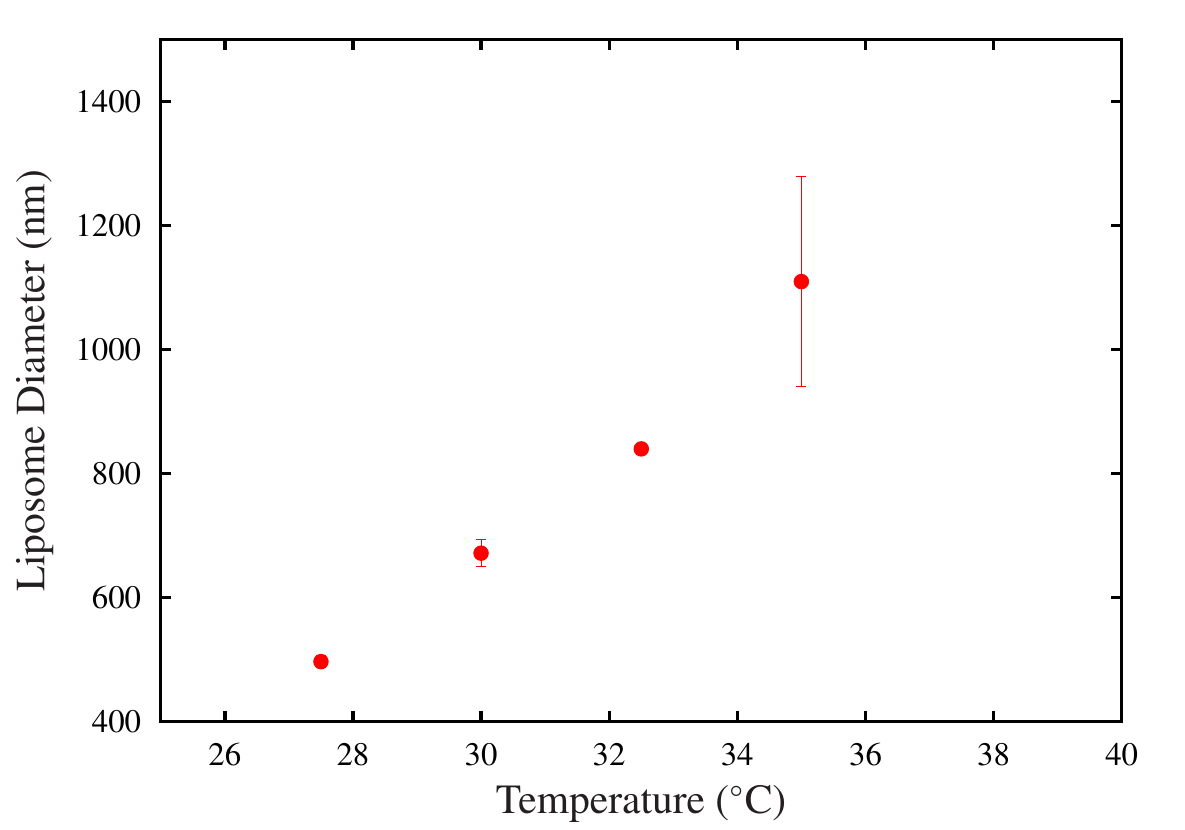}
    \caption{Effect of temperature}
    \label{fig:tempeff}
  \end{subfigure}
  \begin{subfigure}{0.48\textwidth}
    \centering
    \includegraphics[width=\linewidth]{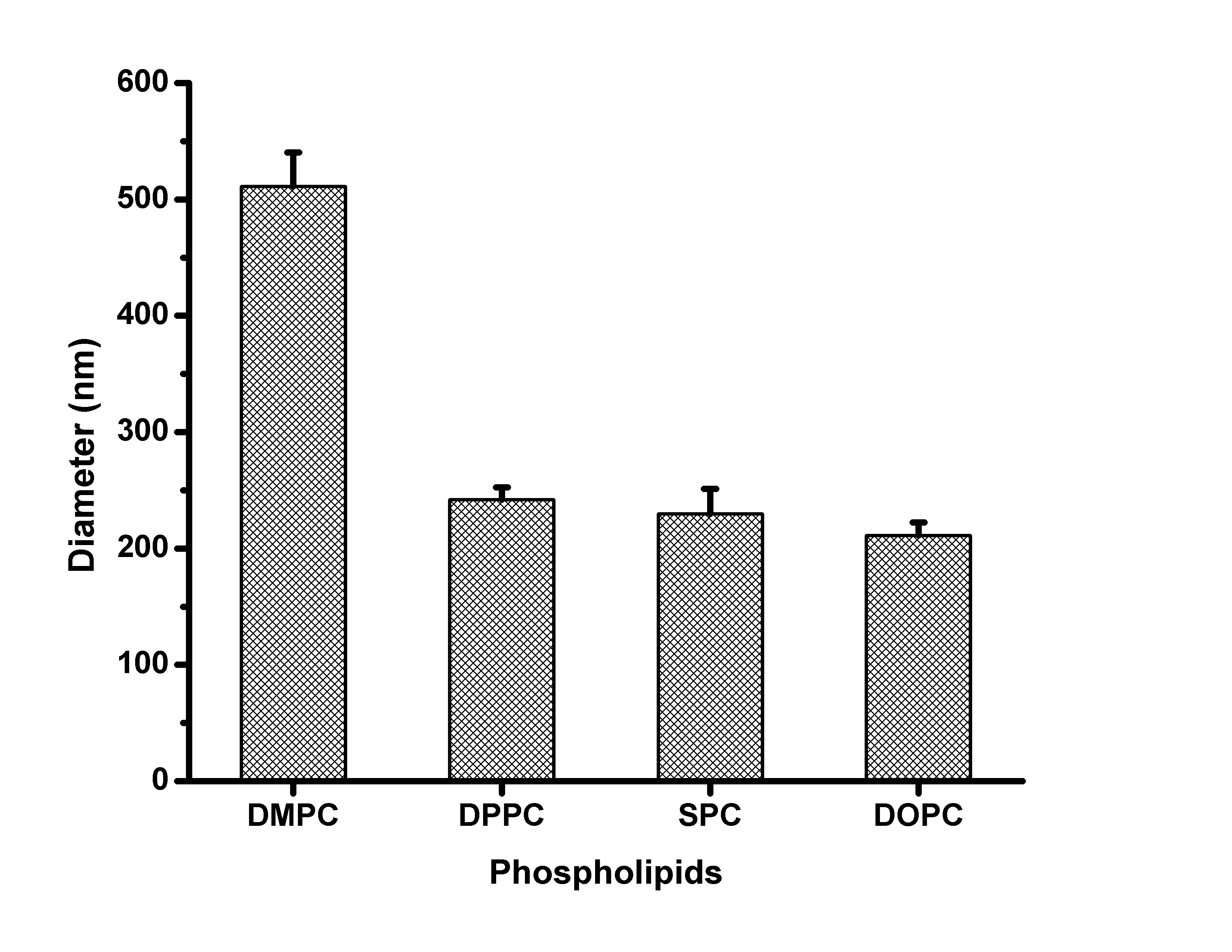}
    \caption{Effect of lipid type}
    \label{fig:lipideff}
  \end{subfigure}

  \caption[Size Variation]{\textbf{Temperature and lipid type
      influences the diameter} (a) Diameter of DMPC liposomes shows an
    increase with increase in temperature.  (b) Liposome size also
    depends on the lipid type, at 27.5\degC. Shown are four lipids
    with various values of acyl chain length:unsaturation: DMPC---14:0,
    DPPC---16:0, Soy PC---mixture-16:0 (14.9\%), 18:0 (3.7\%), 18:1
    (11.4\%), 18:2 (63\%), 18:3 (5.7\%) and DOPC---18:1.}

\end{figure*}

The intrinsic size is expected to change when the modulii $\kappa,
\bar{\kappa}$ or the edge tension $\gamma$ is altered. The temperature
and lipid chemistry (chain length and unsaturation) are two handles
that can readily affect a change in the diameter by this
alteration. Liposomes are formed only at temperatures higher than the
main transition temperature of the phospholipid ($T_\mathrm{m} =
24\degC$ for DMPC), and has a significant effect on the size of
liposomes as seen in \Figref{fig:tempeff}. In the temperature range
studied, the size of the DMPC liposomes shows an increase with
temperature.  In the temperature range 30--40\degC\ the bending
modulus is nearly constant at $\kappa \approx 1.3 \pm 0.1
\tenpow{-19}$~J \cite{Meleard1997}.  However, it is expected that the
edge tension decreases due to a known decrease in the interfacial
tension as the temperature increases.  Lipids of various chain lengths
and unsaturation also form mono-disperse, unilamellar vesicles as seen
in \Figref{fig:lipideff}.  We find that lipids with shorter alkyl
chain length and those with larger number of unsaturated bonds form
larger liposomes, as shown in \Figref{fig:lipideff}, in agreement with
an expected reduction in the bilayer thickness and hence the edge
tension. Accurate measurements of the edge tension values and the
bending modulus for different lipids in the presence of ethanol is
required to further validate our claim.

A remark about the time of synthesis is now in order. In the cuvette
assembly considered here, the vesicle synthesis is completed by about
6 hours \suppl{}. An even faster synthesis time of 15 minutes is
achieved in a horizontal contact of stationary phases inside a
capillary \suppl{}. However, these are not normalised by the amount of
liposomes synthesized. Instead we estimate the time it takes for the
formation of a single vesicle to be about $t_{\mathrm{ves}} \sim
D^{2}/4 \mathcal{D} \approx 5 \tenpow{-5}$~s, for water to diffusively
replace ethanol over the length scale of a vesicle.  Here, $D \approx
500$~nm is the diameter of the vesicle, $\mathcal{D} \approx 1.3
\tenpow{-9}$~m$^{2}/$s is the diffusivity of water in
ethanol. Estimates of the time it takes by other methods is given in
the Supplementary Material \secref{sec:tves}. For a method that
yields monodisperse vesicles by a spontaneous process, this is
possibly the fastest known value, and can therefore be qualified as
``rapid''.

We conclude that SPI mechanism does provide a truly spontaneous
(barrier-free) assembly of lipids to form vesicles, and the observed
diameter of the liposomes is quantitatively explained by Helfrich's
model \citep{Helfrich1974115}.  It is too early to claim that SPI
could be a possible mechanism for the formation of vesicles of a
uniform size under prebiotic conditions, pending identification of
possible surfactants \cite{MonnardDeamer2002}, suitable solvents, and
conditions favouring a interdiffusive mixing. However, it does provide
a direction towards this, grounded in the fact that the there is a
natural length scale dictated by thermodynamics
\citep{Helfrich1974115}.

Nevertheless, the method itself could have fundamental and applied
implications. Due to the stationarity of the SPI method it may now be
possible to infer the dynamics of micelle to vesicle transition in a
single component system using time-resolved scattering, which has been
studied only for kinetically controlled mixed surfactant systems
\citep{EgelSchu1999,Leng2003}.  The proposed mechanism could be
combined with hydrodynamics to understand vesicle size selection in
ethanol injection methods \cite{Phapal2013,Batzri1973,Jahn20042674}.
Our preliminary investigations also reveal that using the SPI process
it is also possible to achieve a high degree of encapsulation (more
than 80\%) of hydrophilic molecules (or drugs) by dissolving it in the
ethanol phase.

\begin{acknowledgments}

  The financial support of Department of Science and Technology,
  Government of India, through a special IRHPA grant, is gratefully
  acknowledged. Facilities in SAIF and CRNTS, IIT Bombay were also
  used in this work. PS also wishes to thank IRCC for the Seed Grant,
  DuPont for the Young Professor fellowship that supported a part of
  this work; colleagues for helpful comments, and KITPC, Chinese
  Academy of Sciences for the Membrane Biophysics Meeting that
  provided an opportunity to discuss this problem at length with the
  several other participants.

\end{acknowledgments}

\bibliography{liposomes-url}

\renewcommand{\appendixname}{Supplementary Material}
\appendix

\section{Thermodynamics of vesicles}
\label{sec:vesthermo}

Here, we briefly recall important results in the thermodynamics of
vesicle formation
\citep{Helfrich1974115,Fromherz1983,Huetal2012,Boal2012}.  The analysis
presented here is restricted to zero-temperature (ignoring entropic
effects). Consider a curved patch (in the form of a spherical cap) of
bilayer membrane (which is also known as bicelle or a bilayer
fragment) of area $A$. There are two important contributions to its
free energy density (per unit area of the patch). First is the elastic bending
energy
\begin{equation}
  \label{eq:gel}
  g_{\mathrm{el}} = ( 2 \kappa + \bar{\kappa} ) \, \left(
    \frac{1}{R} \right)^{2},
\end{equation}
where, $\kappa$ is the elastic bending modulus (splay), $\bar{\kappa}$
is the Gaussian curvature modulus (saddle-splay) and $R$ is the radius
of curvature of the spherical cap. The second contribution is from the
edge of the fragment which has hydrophobic groups of the lipid exposed
to water; the edge energy per unit area of the patch is
\begin{equation}
  \label{eq:ged}
  g_{\mathrm{ed}} = \frac{L \, \gamma}{A},
\end{equation}
where $L$ is the contour length of the edge, and $\gamma$ is the line
tension. The total free energy can be shown to be a function of the
fragment area $A$ and the radius of curvature $R$ \citep{Fromherz1983} as:
\begin{equation}
  \label{eq:Gtot}
  \begin{split}
  G  & = (g_{\mathrm{el}} + g_{\mathrm{ed}}) \, A \\
     & = 
     ( 2 \kappa + \bar{\kappa} ) \, \frac{A}{R^{2}} 
     + 2 \pi \gamma \left( \frac{A}{\pi} - \frac{A^{2}}{4 \pi \,
         R^{2}} \right)^{1/2}.
   \end{split}
 \end{equation}
 The interpretation of this equation is that for a given area $A$ (and
 a lipid-solvent system) the total free energy is a function of
 $R$. The minimum of this free energy determines the likely shape (in
 terms of the mean curvature $1/R$) to find the fragment in.  While
 this form of the energy provides precise expressions for the diameter
 of the spontaneously assembled vesicle, and the free-energy barriers
 for between the two states of a sphere and a disk, a simpler
 derivation is also possible.

 The total energy of a close sphere (without an open hydrophobic edge) is
 \begin{equation}
   \label{eq:Gsph}
   G_{\mathrm{sph}} = 4 \pi ( 2 \kappa + \bar\kappa).
 \end{equation}
 We note that the free energy of a sphere is not dependent on its
 diameter; the free-energy density is dependent on the curvature, but
 this dependence vanishes for the energy itself, when integrated over
 the spherical surface,

 For a flat disk (with a zero curvature) the free energy is
 \begin{equation}
   \label{eq:Gdisk}
   G_{\mathrm{disk}} = L \, \gamma = 2 \pi \, D \, \gamma.
 \end{equation}
 where $D$ is the radius of the disk of the same surface area $\pi
 D^{2}$ of a sphere of diameter $D$.  For a vanishingly small surface
 area ($\pi D^{2}$) the free energy of the disk is lower than that of
 the sphere, and equals it at
 \begin{equation}
   D = 2 \frac{(2 \kappa + \bar \kappa)}{\gamma}.
 \end{equation}
For diameters larger than this, the sphere has a lower free energy;
however, there is an energy barrier for the transition from the disk to
the sphere, which vanishes at
\begin{equation}
  \label{eq:Dintkk}
  D = 4\frac{(2 \kappa + \bar \kappa)}{\gamma}.
\end{equation}
This diameter is therefore the most likely value for the closed vesicle
known as the vesiculation diameter \citep{Fromherz1983}, and called in
the present work as the intrinsic diameter. Earlier works have
neglected the Gaussian curvature modulus
\citep{Fromherz1983,Sackmann1995a}, but recent coarse grained
simulations \citep{Huetal2012} have shown its importance and a way to
obtain it accurately.  It was found that $\bar\kappa/\kappa = - 0.95
\pm 0.1$. Approximating $\bar\kappa \approx - \kappa$, we get from
\Eqref{eq:Dintkk}
\begin{equation}
  \label{eq:Dintk}
  D \approx 4\frac{\kappa}{\gamma}.
\end{equation}

\subsection{Suspension of vesicles}

\begin{figure}[tbp]
  \centering
  \includegraphics[width=\linewidth]{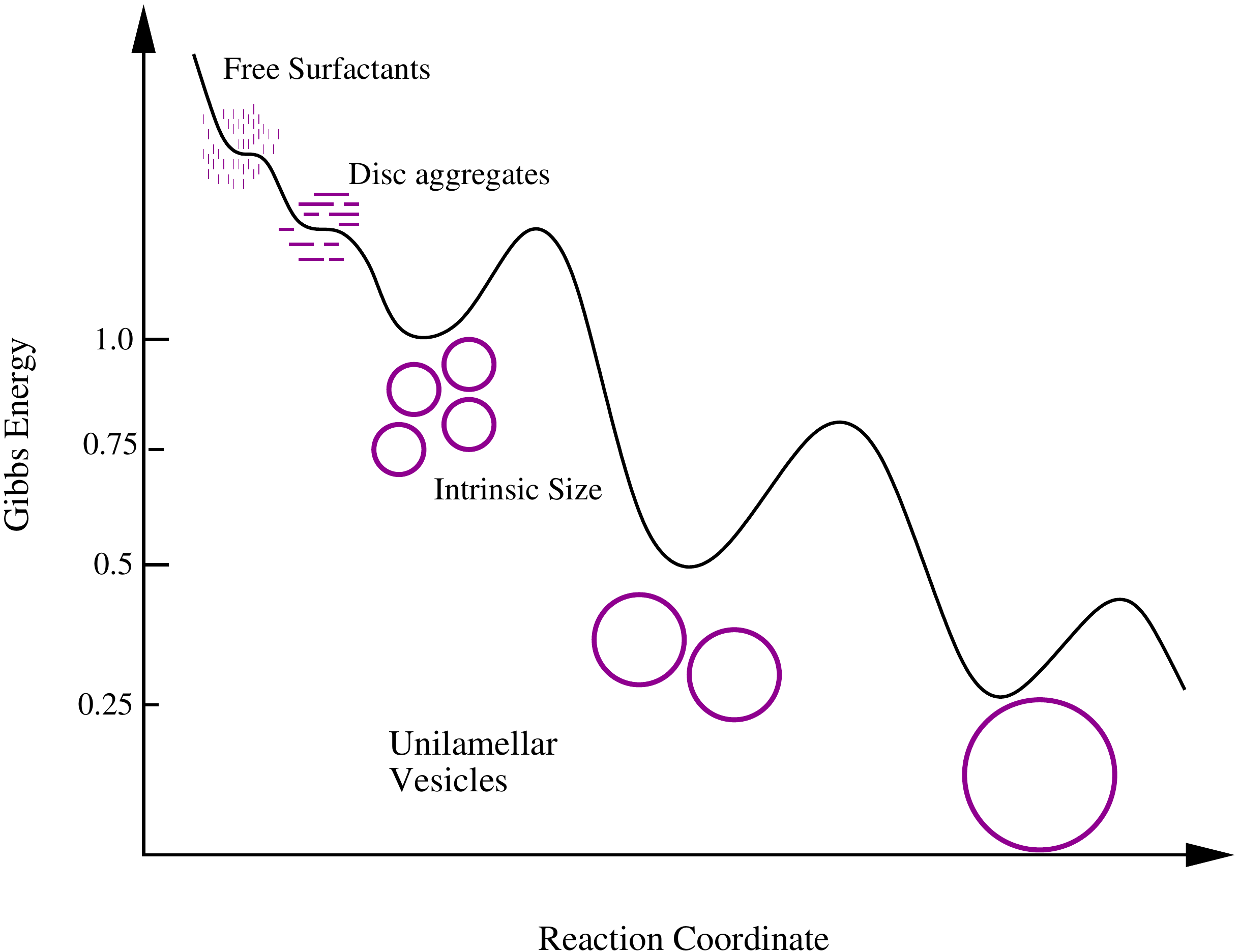}
  \caption{Schematic representation of the free-energy for a
    suspension of vesicles. The highest energy state is that of the free
    surfactants subject to sudden change in the environment that leads
    to a spontaneous aggregation to form discs. The discs, beyond a
    critical diameter, spontaneously close up to form
    vesicles. Vesicles of larger diameter are formed by fusion of
    smaller vesicles reducing the free energy further. But this
    process has an energy barrier of the order of 10~\kT.}
  \label{fig:susen}
\end{figure}

We now consider the free energy of a suspension of vesicles. From
\Eqref{eq:Gsph}, we see that the energy does not depend on the
diameter. Consider a suspension of $M$ number of unilamellar vesicles,
all at its intrinsic diameter $D_{0}$, given by \Eqref{eq:Dintk}.  The
total free energy is therefore $M \, 4 \, \pi \, \kappa$, substituting
the above approximation for $\bar\kappa$. Let the overall energy of
this suspension be normalised to 1 by this value (see
\Figref{fig:susen}).  Now consider the same suspension where every two
vesicles fuse to form one larger vesicle of diameter $D_{1} = \sqrt{2}
\, D_{0}$ (conserving the surface area). The total number of vesicles
would now be $M_{1} = M/2$, and the normalised suspension free energy
would now be $1/2$, since the energy per vesicle is still the
same. This logic can be inductively taken for further fusions to give
$D_{2} = \sqrt{2} D_{1} = 2 D_{0}$ with $M_{2} = M_{1}/2 = M/4$.  This
is schematically illustrated in \Figref{fig:susen}.

From the above analysis it is clear that the intrinsic size is the
highest energy state for vesicles (other higher states for the lipid
solution are the discs and free surfactants, as shown in
\Figref{fig:susen}). These SUVs (small unilamellar vesicles), over
time, will tend towards forming LUVs (large unilamellar vesicles).
However, this transition would need to cross a free energy barrier for
fusion, of the order of 10~\kT\ \citep{Grafmulleretal2007}. Further
lowering of free energy is also possible by formation of
multi-lamellar vesicles, which have an attractive interaction between the
bilayers \citep{Lasic1988} (calculations not shown here).

Therefore, we can conclude that uni-lamellar vesicles are meta-stable
structures kinetically trapped in high energy states.  The kinetic
nature of the vesicular state is unlike other surfactant aggregates
such as micelles which are in a dynamic equilibrium with the dilute
(solution) phase. The vesicle diameter is usually determined by
external parameters from the sample preparation process which leads to
one of these energy traps (by dissipating energy into the system)
\citep{Lasic2001}.  The analysis shown here is minimalistic in that it
does not consider any other shape transformation other than a simple
fusion to conserve area, which is expected to be valid only for dilute
mono-disperse unilamellar suspensions, in the initial times.

\section{Materials and methods}
\label{sec:matmeth}
\paragraph{Preparation of phospholipid solution.}
The phospholipids used to synthesize the liposomes are:
1,2-dimyristoyl-sn-glycero-3-phosphocholine (14:0-DMPC);
1,2-dipalmitoyl-sn-glycero-3-phosphocholine (16:0-DPPC); and
L-$\alpha$-phosphatidylcholine (Soy PC, a mixture of various lipids:
18:2, 16:0, 18:1, 18:3, 18:0 phosphatidylcholine), all purchased
from Avanti Polar Lipids, was used without further
purification. 1,2-dioleoyl-sn-glycero-3-phosphocholine (18:1-DOPC) was
purchased from Sigma Aldrich. Ethanol (AR grade) was purchased from
Merck and MilliQ water was used for the aqueous medium.  All the
lipids from Avanti Polar Lipids were in solution form in chloroform
(20~mg/ml). An appropriate volume of lipid solution is taken out in a
round bottom flask (RBF) and the chloroform is evaporated from the lipid
solution using a rotary evaporator. This forms a thin film on the
glass surface of the RBF. The RBF is then kept in a vacuum desiccator
overnight to remove all the traces of chloroform. An appropriate
volume of ethanol is added to the RBF to make various concentrations
of phospholipid in ethanol (5, 10, 20, 40, 80 mg/ml).

\begin{figure*}[tbp]
  \centering
  \includegraphics[width=0.7\linewidth]{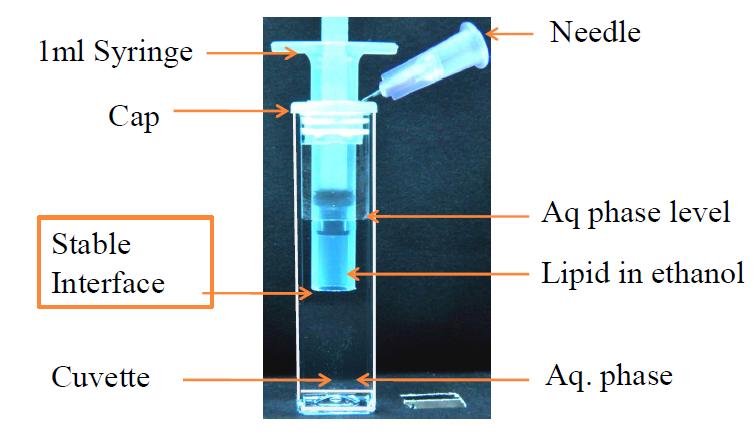}
  \caption[Experimental setup]{\textbf{Experimental setup} Stationary
    phase inter-diffusion apparatus comprises two components---a
    container for aqueous phase (cuvette) and a cut syringe tube
    containing phospholipid-ethanol phase held by the syringe piston.
    The needle is used to gently add the aqueous phase to generate a
    stationary contact with the ethanol phase.}

  \label{fig:exptsetup}
\end{figure*}

\paragraph{Device setup}
The experimental setup is shown in \Figref{fig:exptsetup}. The key
components of the device are: a container for aqueous phase (square
cuvette 10$\times$40~mm) and organic phase container (cut syringe tube
4.8~mm inner diameter). As shown in the setup, a hole is made in the
cuvette cap and the cut tube is inserted through it. The phospholipid
solution in ethanol is filled in the tube (held by the vacuum of the
syringe piston). The aqueous phase is then gently filled in the
cuvette through a syringe needle pierced through the cuvette cap. The
two phases---water and ethanol---are held in stable contact. The
mixing occurs solely due to molecular diffusion of one phase into the
other. Adequate care is taken not to introduce convective mixing in
the two phases near the interface during the process.

\paragraph{Dynamic light scattering (DLS)/Photon correlation
  spectroscopy (PCS)} 
The liposomes were characterized by DLS or PCS (Zetasizer
nano ZS, Malvern, UK) for size and size distribution (polydispersity),
without any further processing, except aqueous dilution, when
required, for reliable measurements.  The laser used internally is
He-Ne laser with 633~nm wavelength and 4~mW power and the scattered
light is detected at a single angle of $173^\circ$. The size and size
distribution was calculated by a software (DTS 6.1, supplied by
Malvern). The polydispersity and average diameter were evaluated using
mono-modal or cumulant analysis method and the intensity-weighted
distribution of diameter was evaluated using multimodal/distribution
analysis. Usual care is taken to prevent dust and other contaminants
in the system.


\paragraph{TEM using Negative staining}
High resolution transmission electron microscopy (Jeol, JEM-2100F) was
used to visualize the liposome lamellarity. Negative staining method
was used to observe the liposomes under the HR-TEM. An electron dense
material---phosphotungstic acid (PTA) 2\%w/v---is used as the negative
staining agent. PTA is acidic in nature and needs to be neutralized;
the pH is adjusted by a dilute NaOH solution. A drop of liposome
solution (10 $\mu$l) is placed carefully on a carbon coated TEM grid
with the help of a micropipette and kept for air drying for 10~minutes,
which is followed by a placing drop of PTA. The sample was air dried
before observing. The operating voltage is kept at 120~kV.

\section{Stages of Liposome formation in SPI}
\label{sec:stages}

\begin{figure*}[tbp]
  \centering
  \includegraphics[width=0.99\linewidth]{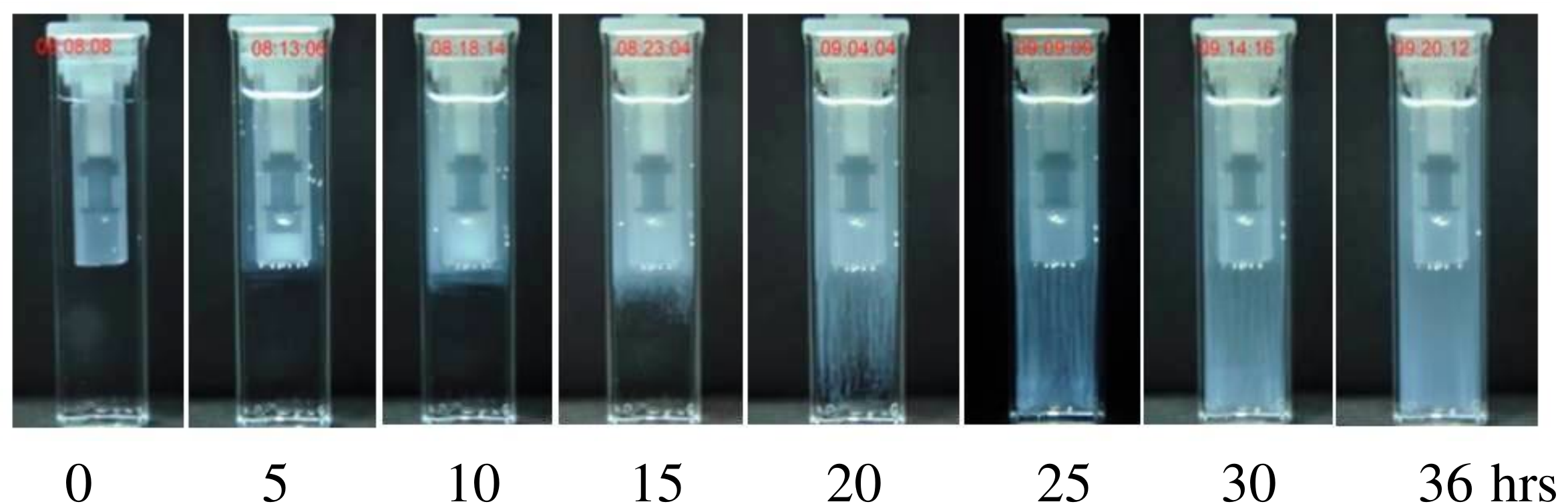}
  \caption[Experimental setup]{\textbf{Stages of liposome formation}
    Time-lapse photographs of liposome formation; each photograph is
    taken in a time interval of 5 hours, showing interdiffusion and
    turbidity of liposome formation. After the initial contact, a
    turbid front moves upward in the ethanol medium, indicating
    liposome formation as water diffuses in. At around 6 hours the
    entire section of the phase inside the cut-syringe becomes
    uniformly turbid. After about 15 hours, strings of suspension move
    downwards, which laterally diffuse to form a uniform suspension by
    about 36 hours.  The time stamp is labelled at the top of each
    frame in the format DD:MM:HH (denoting day, minute, and hour
    respectively).}
  \label{fig:timelapse}
\end{figure*}

The phospholipid dissolved in ethanol is taken in an inverted tube and
is in contact with an aqueous phase (cuvette) as shown in
\Figref{fig:exptsetup}.  As soon as the interface contact is
established, the progress of liposome formation is clearly visible in
the form of a turbid front that moves upward through the ethanol
section.  Time lapse photographs of the visual changes observed are
shown in \Figref{fig:timelapse}.  

A copy of the time-lapse video of the formation of liposome suspension
is available from:
\begin{quote}
\url{http://goo.gl/KOW3VN}
\end{quote}
The video has been made from photographs taken in 10~min time
intervals (the timestamp is labelled on the top of the cuvette, in the
format DD:HH:MM). Three distinct phenomena are observed---(a)
Initially, both the lipid and aqueous phases are clear, and soon
after, a turbid layer is observed inside the inverted tube which moves
upwards, indicating that liposomes are being formed. This is completed
by about 6~hours (b) As the white region moves vertically, some amount
of liposome suspension also comes out in the aqueous phase (cuvette)
and goes up towards the water-air interface, due to the lower density
of ethanol that is diffusing into the aqueous phase. This is followed
by a downward movement of ``strings'' of suspension towards the bottom
of the cuvette (at a steady velocity of about 1~$\mu$m per~sec). (c)
Finally the strings diffuse out to form a uniformly turbid milky white
suspension.

The time of 6~hours for the first stage roughly agrees with the time
scale of diffusion of water through ethanol through a distance of
1~cm, the length of the ethanol section in the cut-syringe (see
\secref{sec:tves}).  In the second stage, we believe, the strings
form because of a viscous fingering instability occurring due to a
mean motion generated by diffusiophoresis of the liposomes in a
gradient of ethanol. Further experiments and analyses are underway to
confirm this claim.

\section{Estimate of the Intrinsic Diameter}
\label{sec:dint}

The intrinsic diameter of the vesicle is given by \Eqref{eq:Dintk}. In
the SPI method, the hydrodynamic diameter of the liposomes was found
to be 530~nm for the DMPC lipids.  Our claim that this is the
intrinsic diameter can be validated by using independent measurements.

The line tension $\gamma$ can be estimated by a molecular geometry
model of the of arrangement of lipids along the edge
\citep{LyLongo2004}:
\begin{equation}
  \label{eq:gammalylongo}
  \gamma = \frac{5 \pi}{16} \, \sigma \, b_{0} \approx \sigma \,
  b_{0},
\end{equation}
where $\sigma$ is the interfacial tension between the hydrophobic tail
and the surrounding solvent, and $b_{0}$ is the length of the
hydrophobic tail of the lipid. For DMPC lipids the length of the
hydrophobic tail $b_{0} = 2.54/2 \approx 1.27$~nm \citep{Kucerka2005}.
We note that recent measurements of the line tension by observing the
dynamics of pore closure \citep{Karatekin2003}, find the line tension
to be dependent on the manufacturing source (possibly because of small
amounts of impurities). In the absence of measurements of $\gamma$ for
our sample, we choose to use the above theoretical estimate
consistently for the two cases below.

\subsection{Intrinsic diameter in water}
When the surrounding solvent is water, we use an water-oil interfacial
tension of $\sigma = 35$~mN/m. This gives the line tension to be
$\gamma_{\mathrm{w}} = 44$~pN.  For DMPC bilayer in water the bending
modulus is $\kappa = 1.5 \pm 0.06 \tenpow{-19}$~J at 27\degC\
\citep{Meleard1997}. 
 Substituting these estimates we find
$D_{\mathrm{water}} = 13$~nm. Experimentally, it is observed that more
than two hours of sonication results in a diameter of 17~$\pm 6$~nm
\citep{Maulucci2005}.  

\subsection{Intrinsic diameter in ethanol-water mixture}

In the method employed in \secref{sec:matmeth} the self-assembly takes
place in the presence of ethanol, and the estimates of the physical
properties and hence the diameter will differ.  The bending modulus of
lipid membranes is reduced in the presence of ethanol by about 30\% at
20\%~v/v of ethanol for SOPC membrane vesicles
\citep{LyLongo2004}. This implies for DMPC we can expect $\kappa
\approx 1 \tenpow{-19}$~J in the presence of ethanol.  More
significant is the reduction in the line tension alcohol can induce to
an oil water interface. The lowest measured interfacial tension for
this system is about $\sigma \approx$~1~mN/m
\citep{Sanaiotti2010}. Therefore the line tension can be estimated
from \Eqref{eq:gammalylongo} as $\gamma \approx 1.3$~pN.  This gives
an estimated vesicle diameter of 300~nm, which agrees to within an
order of magnitude of the experimentally observed value of 530~nm, and
is reasonable given approximate nature of the property estimation we
have used.  More accurate measurements of the either the interfacial
or the edge tension \citep{Karatekin2003} for various lipids in the
presence of ethanol may confirm our hypothesis of the intrinsic size.

\section{``Minimum'' vs ``Intrinsic'' diameter}
\label{sec:min-int}

It has been customary to refer to the diameter given in
\Eqref{eq:Dintk} as the minimum diameter of vesicles, because upon
subjecting to external driving (sonication or turbulence) the average
diameter of liposomes in an aqueous lipid solution will be given by
\Eqref{eq:Dintk}; also see \Figref{fig:susen}. The energy dissipated into
the system by external means, which can break larger and multilamellar
vesicles, leads to the reassembly by aggregation (of discs and free
lipids).  The qualifier `minimum' is used because vesicles of a
diameter smaller than the diameter given by \Eqref{eq:Dintk} is highly
improbable, owing to the high cost of bending a disc of small
diameters.

However, we choose to call it as the `intrinsic' diameter since it is
more generic.  In the present system the observed diameter of 530~nm
is intrinsic for the lipid+ethanol+water system.  It is the `minimum'
only in the surroundings where it forms (in the presence of a
significant concentration of ethanol (25--50\%) in water).  However,
as water completely replaces ethanol, the overall concentration of
ethanol in the final surrounding can be very low (less than 1\%
depending on the total volumes of mixtures taken initially). In such a
situation, the minimum diameter obtained by sonication (of aqueous
suspension of lipids) would be much smaller, around 20~nm (see
\secref{sec:dint}).  It would thus be misleading to call a 500~nm
vesicle to be in its minimum diameter.  The nomenclature of
`intrinsic' vs `minimum' is only one of semantics. Using `intrinsic'
also connotes a physical basis to a spontaneously formed vesicle,
whereas `minimum' can also refer to something achieved by external
force.

\section{Capillary Setup for Stationary Phase Inter-diffusion}
\label{sec:capil}

\begin{figure*}[tbp]
  \centering
  \includegraphics[width=0.8\textwidth]{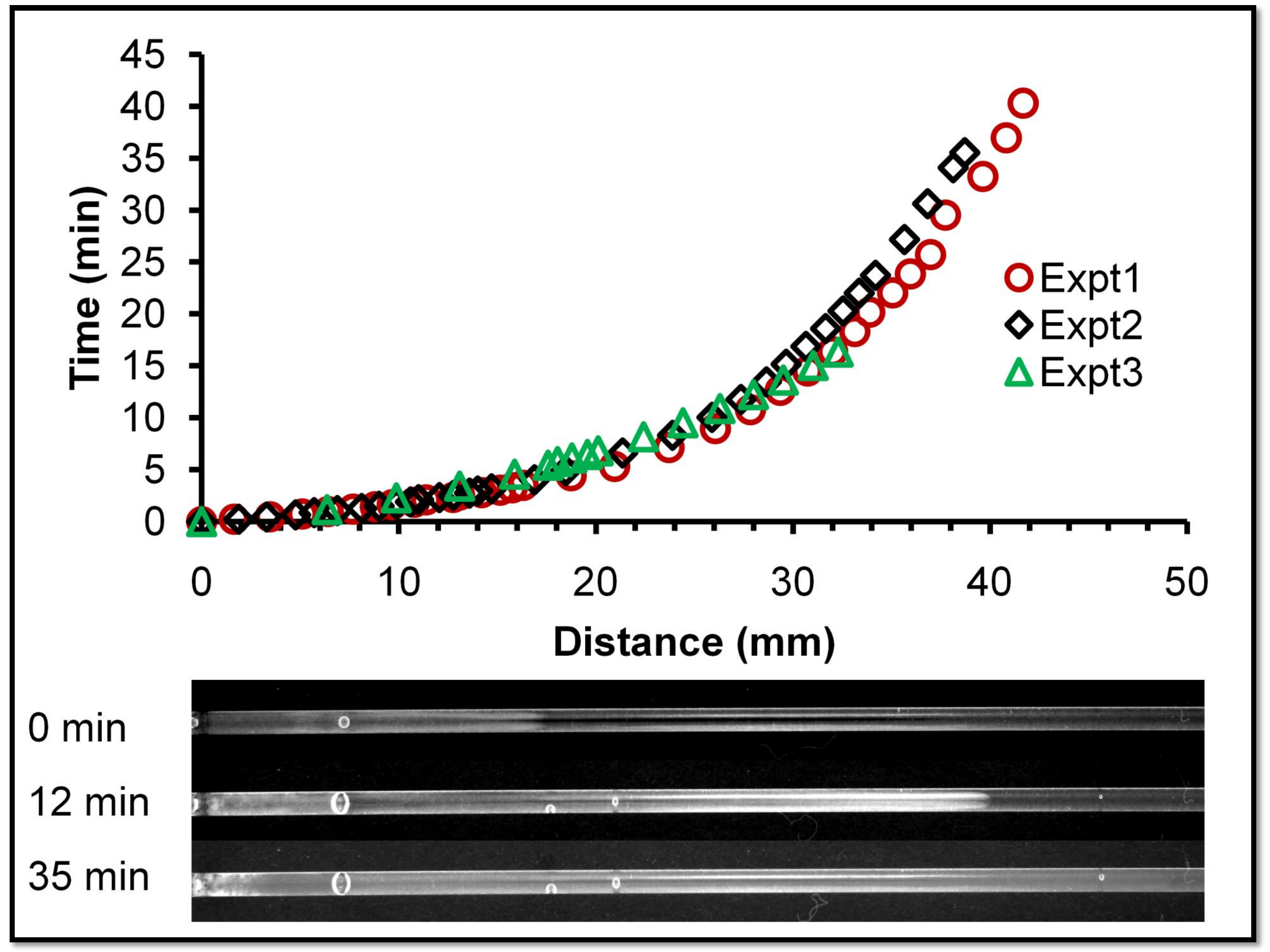}
  \caption[Capillary SPI]{\textbf{Capillary setup of SPI mechanism.}
    Liposome synthesis in a capillary by the SPI method, shows
    completion of liposome formation in less than 45 minutes. Here the
    distance moved by the turbid front is plotted against time. Also
    shown below are sample images of the vesicle dispersion forming
    inside the glass capillary at different times.}
  \label{fig:capsynth}
\end{figure*}

The cuvette and syringe method (\Figref{fig:exptsetup}) takes $\sim$6
hours to form the liposomes (1~cm height and 4.8 mm diameter tube).
The formation time is observed to be considerably smaller when
capillaries are used. Glass capillaries 10~cm long and 1~mm in inner
diameter is used with one end sealed. The procedure in short is---the
aqueous phase is filled up to 8~cm with the help of a glass syringe
attached to a 12~cm long 30~G SS needle; The phospholipid-ethanol
solution is taken in another syringe and filled up to 1.5~cm by
keeping a 1/2~cm air gap between the phases. Then the two miscible
solutions are brought in contact by sucking out the air bubble through
another needle. Care is taken to avoid any convective mixing at the
interface otherwise a multi-modal size distribution of liposomes is
observed. The total time required to form liposomes is less than an
45~minutes as seen in \Figref{fig:capsynth}.  This can be
significantly reduced even less to about 15~minutes by gently tapping
the capillary, to release vapour bubbles that occlude the motion of
the turbid front.  We believe the reason for the faster synthesis in a
capillary setup is due to a phoretic motion of the liposomes driven by
a concentration gradient of ethanol in water (diffusiophoresis,
\cite{Brady2011}) or because of a viscosity gradient, which in turn
leads to faster mixing of water in ethanol in the opposite direction,
furthering the assembly of liposomes.

\section{Time for vesicle formation}
\label{sec:tves}

As shown in \secref{sec:stages}, the entire process in the cuvette
cell configuration takes about 36 hours. However, samples taken as
soon as the turbid front reaches the top of the syringe section at
about 6 hours, also show the same monodisperse distribution of
liposomes, implying the completion of vesicle synthesis. 

The time it takes for water to diffusively transport along the length
of the section inside the syringe can be estimated to be
$t_{\mathrm{water}} \sim L^{2}/4 \mathcal{D} \approx 5$~hrs, where
$L=1$~cm section that initially contains the lipid solution in
ethanol, and $\mathcal{D} \approx 1.3\tenpow{-9}$~m$^{2}$/s is the
diffusivity of water in ethanol.  This is in near agreement with the
above observation on completion of liposome synthesis, and in line
with the hypothesis of the diffusive mechanism proposed (main
text). Though the 6 hours in the cuvette setup and 15~minutes in the
capillary setup (\secref{sec:capil}) seems fast for a method that yields
monodisperse liposomes in a single step, this needs additional
qualification.

We now provide detailed comparison of the time it takes for liposome
synthesis in this method with other methods.  It will be inappropriate
to simply compare the overall time of an experiment, because this also
involves the total amount of liposomes formed, which could vary from
one method to another, given that there has not been any standardised
study. Instead it is plausible to compare the expected time for
formation of a single vesicle in each of the methods.
\Tabref{tab:tmeth} lists the order of magnitude time taken in various
methods. The estimates have been obtained by using the time spent only
in the actual liposome formation step. For the present method, we use
the approximate time it takes for water to diffusively replace ethanol
over a length scale of a liposome diameter: $t_{\mathrm{SPI}} \sim
D^{2}/4 \mathcal{D} \approx 5 \tenpow{-5}$~s.  where $D \approx
500$~nm is the liposome diameter, and $\mathcal{D} \approx 1.3
\tenpow{-9}$~m$^{2}$/s is the diffusivity of water in ethanol.  This
shows that a liposome is formed in about 50~microseconds.

For methods which use shear forces to drive assembly (or break up
liposomes), there are no detailed studies of the time taken; in lieu
of which, we have used a typical turbulent (Kolmogorov) time scale
$t_{\mathrm{turb}} \sim \Rey^{-1/2} d/U \approx 1 \tenpow{-5}~$s for a Reynolds
number \Rey\ of 10000, generated in a macroscopic length of $d=1$~mm
with a bulk fluid velocity $U\approx 1$~m/s.

It can be concluded from analysing \Tabref{tab:tmeth} that the time
taken for the synthesis of a single vesicle in SPI is among the
fastest in comparison with other methods that result in monodisperse
distribution by a spontaneous process. Even in comparison with the
other externally driven methods the SPI method is comparable in the
order of magnitude of the time taken. We can also infer that since the
SPI method is interface driven, a larger interfacial contact will
produce a greater yield of liposomes in the same time for a given
initial concentration of lipids.

\begin{table*}[tbp]
  \caption{Estimate of order of magnitude time taken for the formation
    of a single vesicle in various methods against the size distribution
    and the driving force for size selection.}
\label{tab:tmeth}
\begin{tabular}{p{0.2\linewidth}p{0.2\linewidth}p{0.2\linewidth}p{0.2\linewidth}}
\hline
Method                                   & Distribution      & Driving Force for size selection     & Time Taken one liposome (order of magnitude) \\
\hline
Ethanol injection                        & Polydisperse      & Fluid shear                          & 10 microsecond                               \\
Reverse phase evaporation                & Polydisperse      & Fluid shear (mixing upon sonication) & Minute                                       \\
Detergent depletion (extemporaneous)     & Monodisperse      & Rate of detergent removal            & Minute                                       \\
Electroformation                         & Monodisperse      & Electric field                       & Minute                                       \\
Sonication of aqueous solution of lipids & Monodisperse      & Fluid shear                          & 10 microsecond                                  \\
French press                             & Monodisperse      & Fluid shear                          & 10 microsecond                               \\
Thin film hydration                                & Polydisperse, MLV & Spontaneous (or gentle shear)        & Hour                                         \\
Bilayer hydration                      & Monodisperse      & Spontaneous                          & Minute                                       \\
SPI                                      & Monodisperse      &
Spontaneous                          & 10 microsecond
\\
\hline
\end{tabular}
\end{table*}

\section{Concentration of Liposome Suspension}
\label{sec:conc}

As shown in the main communication, in the SPI process the
concentration of lipids does not influence the size of liposomes.
Increasing the concentration of lipids only results in more number of
liposomes formed. We observe a suspension of a very high liposome
fraction, as visually seen by a highly turbid mass inside the
cut-syringe section.  However, we were unable to exactly quantify the
concentration; as shown in \Figref{fig:timelapse}, a part of the
liposome suspension comes out of the cut syringe section as the turbid
front moves upwards.  This makes it difficult obtain accurate
measurements.  However, an estimate of the volume fraction of the
liposome suspension can be obtained based on the average liposome
diameter and the amount of lipids used.

Assuming that all of the lipids dissolved in ethanol form liposomes of
the same size, and all of them are contained in the cut-syringe
section, we can estimate the volume fraction of the liposome
suspension in the cut-syringe section:
\begin{equation}
  \phi = \frac{c\, N_{\mathrm{A}}\, D \, a_{0} }{12 M}.
\end{equation}
Here, $c$ is the mass concentration of lipids, $N_{\mathrm{A}}$ is the
Avogadro number, $D$ is the average diameter of the liposomes, $a_{0}$
is the area per head group of lipid, and $M$ is the molecular weight
of lipid.

For DMPC lipid, $M=678$~g/mol, and $a_{0} = 0.71$~nm$^{2}$.  From the
DLS measurements we obtain the diameter of liposomes $D\approx
500$nm. For a concentration of lipids at $c=20$~mg/ml, we obtain the
volume fraction to be
\begin{equation}
  \phi \approx 0.5
\end{equation}
which is close to the maximum volume fraction of $\phi = 0.64$ for
random packing of spheres.  Higher concentrations lead to even higher
volume fractions. For the highest used concentration, $c=80$mg/ml, we
get $\phi \approx 2$, which is higher than the close packing
fraction. In this case, as mentioned above, the liposome suspension no
longer remains entirely confined to the cut-syringe section, and the
estimate is not a correct reflection of the actual concentration.

A high concentration liposome suspension can be harvested by carefully
removing the cut-syringe assembly from the cuvette, as soon as the
turbid front reaches the top of the cut-syringe.  As indicated in the
main communication, the liposome formation is completed by this time
(at about six hours) and a mono-disperse distribution is obtained even
without waiting for 36~hours for the entire process to complete.

\end{document}